\documentclass[12pt]{iopart}

\usepackage{graphicx}

\newcommand{\ud}{\mathrm{d}}

\begin{document}

\title[]{Statistical hadronization model predictions for charmed hadrons at LHC}

\author{A Andronic$^1$, P Braun-Munzinger$^{1,2}$, K Redlich$^3$, J Stachel$^4$}
\address{$^1$ Gesellschaft f\"ur Schwerionenforschung, GSI, 
D-64291 Darmstadt, Germany}
\address{$^2$ Technical University Darmstadt, D-64283 Darmstadt, 
Germany}
\address{$^3$ Institute of Theoretical Physics, University of 
Wroc\l aw, PL-50204 Wroc\l aw, Poland}
\address{$^4$ Physikalisches Institut der Universit\"at Heidelberg,
D-69120 Heidelberg, Germany}

\begin{abstract}
We present predictions of the statistical hadronization model for charmed 
hadrons production in Pb+Pb collisions at LHC.
\end{abstract}

The results presented below are discussed in detail in our recent 
publication \cite{Andronic:2006ky}. 
We summarize here the values of the model parameters:
i) characteristics at chemical freeze-out: temperature, $T$=161$\pm$4 MeV;
baryochemical potential, $\mu_b$=0.8$^{+1.2}_{-0.6}$ MeV; volume 
corresponding to one unit of rapidity $V$=6200 fm$^3$;
ii) charm production cross section:
$\ud \sigma_{c\bar{c}}^{pp}/\ud y=0.64^{+0.64}_{-0.32}$ mb.

\begin{figure}[htb]
\begin{tabular}{lr}
\begin{minipage}{.49\textwidth}
\vspace{-1cm}
\centering\includegraphics[width=1.15\textwidth]{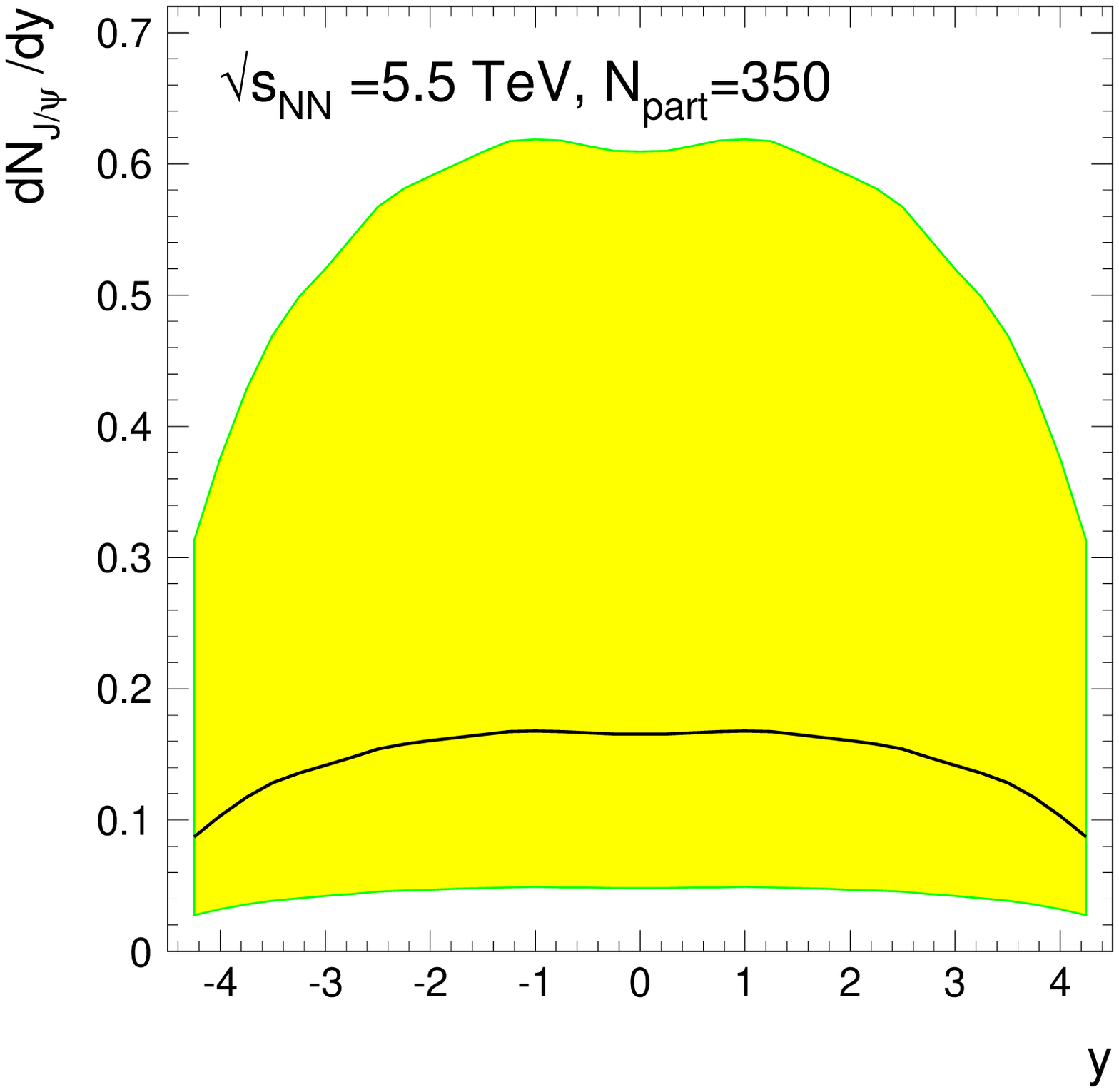}
\end{minipage}  & \begin{minipage}{.49\textwidth}
\vspace{-1cm}
\centering\includegraphics[width=1.15\textwidth]{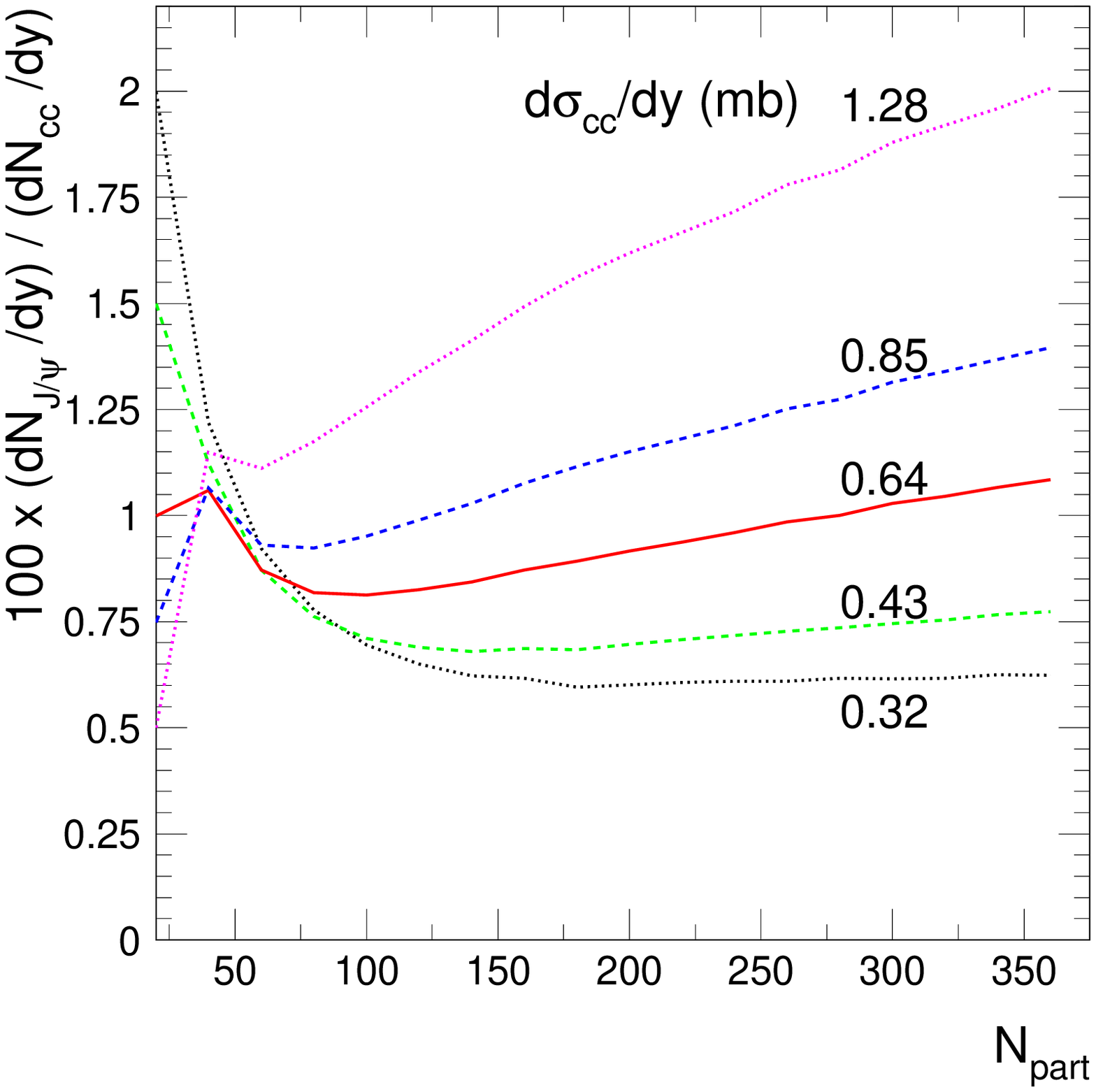}
\end{minipage} \end{tabular}
\vspace{-.8cm}
\caption{Predictions for $J/\psi$ yield: rapidity distribution for 
central collisions (left panel) and centrality dependence 
of the yield relative to the charm production yield for different
values of the charm cross section indicated on the curves (right panel).}
\label{aa_fig1c}
\end{figure} 

In Fig.~\ref{aa_fig1c} we present predictions for the yield of $J/\psi$.
The left panel shows the rapidity distribution with the band reflecting
the uncertainty in the charm production cross section.
The right panel shows the centrality dependence of the yield relative 
to the charm production yield for five values of the input charm cross section.

The statistical hadronization model predictions for charmed hadron yield 
ratios in central Pb+Pb collisions at LHC are shown in Table~\ref{tab_aa_t1}.
We expect that these ratios are independent of centrality down to values
of $N_{part}\simeq$100.

\begin{table}[hbt]
\caption{Predictions of the statistical hadronization model for charmed 
hadron ratios for Pb+Pb collisions at LHC. The numbers in parantheses 
represent the error in the last digit(s) due to the uncertainty of $T$.\\}
\label{tab_aa_t1}
\begin{tabular}{ccccccc}
$D^-/D^+$ & $\bar{D_0}/D_0$ & $D^{*-}/D^{*+}$ & $D_s^-/D_s^+$ & 
$\bar{\Lambda_c}/\Lambda_c$ & $D^+/D_0$ & $D^{*+}/D_0$\\ \br
1.00(0) & 1.01(0) & 1.01(
0) & 1.00(1) & 1.00(1) & 0.425(18) & 0.387(15) \\ \br \\
$D_s^+/D_0$ & $\Lambda_c/D_0$ & $\psi'/\psi$ & $\eta_c/\psi$ & 
$\chi_{c1}/\psi$ & $\chi_{c2}/\psi$ &\\ \br
0.349(14) & 0.163(16) & 0.031(3) & 0.617(14) & 0.086(5) & 0.110(8) & \\ \br 

\end{tabular}
\end{table}

\vspace{-1.5cm}
\begin{figure}[htb]
\begin{tabular}{lr} \begin{minipage}{.49\textwidth}
\centering\includegraphics[width=1.22\textwidth]{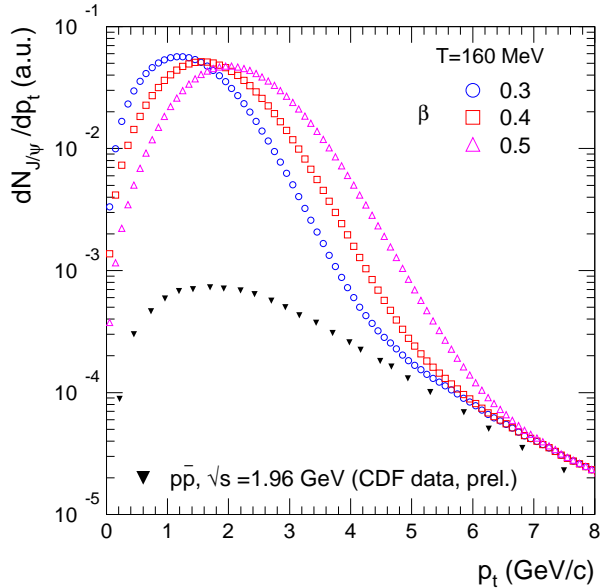}
\end{minipage}  & \begin{minipage}{.45\textwidth}
\vspace{-.8cm}
\caption{Predictions for momentum spectrum of $J/\psi$ meson
for different values of the average expansion velocity, 
$\beta$, for central Pb+Pb collisions ($N_{part}$=350). 
Also included is the measured spectrum in p$\bar{\mathrm{p}}$ collisions at 
Tevatron \cite{Pauletta:2005nt}, which is used to calculate the contribution 
from the corona (see ref. \cite{Andronic:2006ky}).}
\label{aa_fig2c}
\end{minipage} \end{tabular}
\end{figure}

Following from our model assumption of charm quark thermalization and 
assuming decoupling of charm at hadronization, the transverse momentum 
spectra of charmed hadrons can be calculated \cite{Andronic:2006ky}. 
As seen in Fig.~\ref{aa_fig2c}, a precision measurement of the spectrum 
of $J/\psi$ meson will allow the determination of the expansion velocity 
in QGP.

\section*{References}
\bibliographystyle{unsrt}
\bibliography{charm}


\end{document}